# Optic Fingerprint(OFP): Enhancing Security in Li-Fi Networks

Ziqi Liu[§], Xuanbang Chen[§†], Xun Zhang[§],
[§]Institut Supérieur D'électronique de Paris, France, [†]Nanchang University, China,
Email: *ziqi.liu@isep.fr, xuanbang.chen@ext.isep.fr, xun.zhang@isep.fr*

*Abstract*—This paper explores the integration of hardware device fingerprint extraction and identification processes into the LiFi standard protocol. The primary objective is to enhance the communication security of LiFi networks by leveraging unique hardware characteristics as a layer of authentication and verification. The Optic Fingerprint (OFP) model leverages the inherent nonlinear characteristics of Light-Emitting Diode (LED) devices, extracting amplitude-based feature vectors with high precision. Unlike traditional Radio Frequency (RF) fingerprinting, these features are designed specifically for optical wireless communication, capturing LED-specific variations in both the time and frequency domains. While maintaining compliance with the IEEE 802.15.7 LiFi standard requirements, extensive experimental validation demonstrates the framework's robustness, achieving up to 90.36% classification accuracy across 39 white LEDs under varying Signal-to-Noise Ratios (SNR: 10–30 dB).

## I. INTRODUCTION

LIGHT-FIDELITY (LiFi) technology is emerging as a key solution for high-speed, interference-free wireless communication in dense Internet of Things (IoT) environments [1]. However, its broadcast nature makes it vulnerable to security threats, such as eavesdropping and unauthorized access [2]. As next-generation networks emerge, traditional cryptographic methods outlined in IEEE 802.15.7 [3], [4] face challenges such as complex key management [5], risks of key exposure [6], and vulnerabilities to quantum attacks [7]. These challenges necessitate robust, scalable security mechanisms for future LiFi systems.

Device Fingerprinting (DF) provides a promising solution by leveraging hardware-specific variations to create unique, non-replicable identifiers [8], [9]. DF eliminates reliance on pre-shared keys, resists computational attacks, and supports real-time authentication with minimal overhead [10], [11]. In Radio Frequency (RF)-based systems, such as Wireless Fidelity (Wi-Fi), Long Term Evolution (LTE), and Long Range (LoRa), DF has been effectively applied to enhance both device and protocol security. For instance, Peng et al. embedded RF fingerprints into LTE protocols to secure cellular uplink channels, demonstrating robustness against channel variations and noise [12]. Similarly, Shen et al. proposed scalable DF methods for LoRa networks, enabling robust authentication even in resource-constrained environments by considering protocol limitations during fingerprint design [13]. These efforts illustrate how DF can be tailored to specific communication protocols to improve security and scalability. Those works demonstrate the importance and feasibility of DF in network security. Meanwhile, LiFi networks are gaining increasing attention, with the integration of LiFi and 5G validated in [14], [15]. Optical Wireless Communication (OWC) networks inherently possess robust security advantages. This paper investigates how Optic Fingerprint (OFP) can be leveraged to enhance the security of LiFi networks. Furthermore, our proposal fully complies with the IEEE 802.15.7 standard.

To date, most research on OFP has primarily focused on device modeling and ensuring its stability. Shi et al. explored frequency response measurements, notably the Scattering Parameter $S_{21}$ (S21) as Light-Emitting Diode (LED) fingerprints [16], [17]. Liu et al. proposed a deep feature separation network for emitter fingerprint identification [18].) These methods rely on features (e.g., LED emission spectra [19] or S21 parameters [16], [17]) that are typically inaccessible during real-time communication, highlighting the necessity of integrating OFP mechanisms directly into existing communication networks to enhance their security and practicality.

In this context, we propose an OFP-based identification framework designed specifically in compliance with the IEEE 802.15.7 LiFi standard. The primary contributions are summarized as follows:

- A novel method using amplitude-based features (Mean Amplitude, Variance, Baseband Amplitude) tailored for OWC, ensuring robust LED differentiation.
- An OFP framework integrating an IEEE 802.15.7-compatible data frame, feature extraction, and machine learning classifiers for scalable, real-time identification.
- Validation achieving 90.36% classification accuracy across 39 LEDs under diverse conditions (Signal-to-Noise Ratio (SNR): 10–30 dB, 1–30 cm) and Bit Error Rate (BER) compliance with IEEE 802.15.7 standards in lab environment.

The rest of this paper is organized as follows. Section II analyzes the design and integration of the OFP framework into LiFi networks, focusing on feature generation and its compatibility with the OWC link. Section III presents experimental results and simulations to evaluate the performance and robustness of the OFP framework under varying conditions.

The authors gratefully acknowledge the financial support of the Chinese Scholarship Council and the EU Horizon 2020 program towards the 6G BRAINS project H2020-ICT 101017226. (Corresponding author: Xun ZHANG.)

Section IV concludes the paper and discusses future research directions.

## II. OPTIC FINGERPRINT DESIGN

### A. OFP definition

LiFi networks inherently operate as bidirectional systems to enable uplink and downlink communication. However, to simplify the analysis and focus on the essential aspects of OFP extraction, this paper considers only the unidirectional OWC link. This simplification allows us to analyze better the OFP originating from LED hardware variations and their propagation through the OWC link.

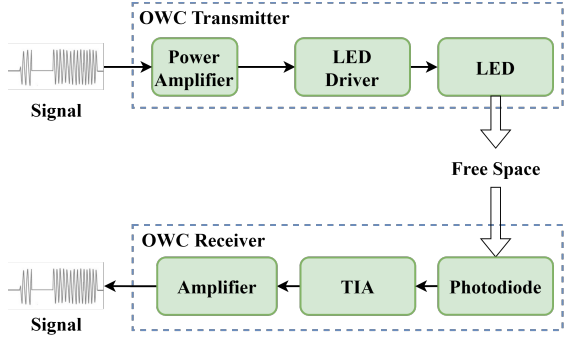

Fig. 1: OWC downlink diagram.

OFP in previous research has primarily been attributed to LED hardware-specific variations. These intrinsic differences are modeled using $H_{\text{led}}(t)$, which captures the nonlinear and device-specific properties of LEDs [20]. The LED model is expressed as:

$$H_{\text{led}}(t) = \frac{1}{2\pi} \int_{-\infty}^{\infty} \frac{\Gamma \cdot \eta_{ph} \beta_{sp} Z_E Z_{ph} \cdot \sum_{i=1}^{m} \frac{1}{r_{iqr}}}{Z_E + R_s + Z_s} e^{-j\omega t} \, d\omega, \quad (1)$$

where $Z_s$ represents the output impedance of the signal source. The terms $Z_E$ and $Z_{ph}$ denote the equivalent impedances of the electrical and optical loops, while $\Gamma$ accounts for attenuation from the coated phosphor. The photon emission efficiency of the LED is denoted as $\eta_{ph}$, and $R_s$ represents the parasitic resistance of the LED package. The differential resistance is denoted as $r_{iqr}$, and $\beta_{sp}$ represents the spontaneous emission coefficient.

These parameters reflect the intrinsic differences in LED hardware, which lead to unique signal distortions. The variations in $H_{\text{led}}(t)$ can be visualized in Fig. 2, highlighting device-specific impacts on the transmitted signal.

To evaluate the impact of the subtle hardware differences for OFP in the context of the OWC link, it is coupled with a typical LED model $h_{\text{led}}$ [20], a typical OWC channel model $h_{\text{channel}}$ [21] and linear response of the OWC receiver $h_{rx}$, forming an OWC transmission link model $h_{\text{owc}}$, and is expressed as :

$$H_{\text{owc}}(t) = H_{\text{led}}(t) \otimes H_{\text{channel}}(t) \otimes H_{\text{rx}}(t), \quad (2)$$

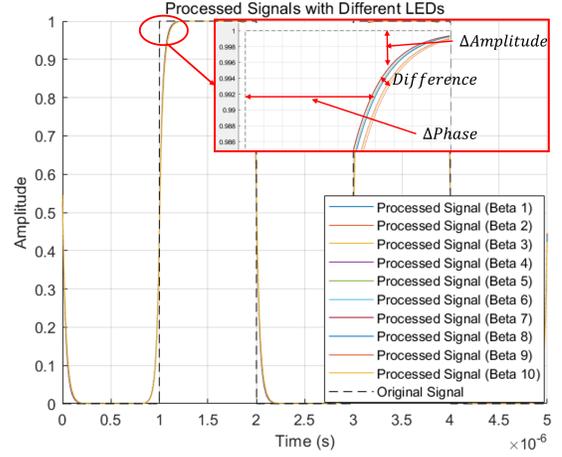

Fig. 2: Comparison of Transmitted and Processed Signals: Differentiation Across Various LEDs.

where $H_{\text{channel}}(t)$ and $H_{\text{rx}}(t)$ are modeled as:

$$H_{\text{channel}}(t) = \delta(t) \cdot \frac{(\mu + 1) A_{\text{rec}} \cos(\psi)^\mu \cos(\theta)}{2\pi D^2}, \quad (3)$$

$$H_{\text{rx}}(t) = \delta(t) G_{\text{am}} \int_{\lambda_{\min}}^{\lambda_{\max}} \kappa(\lambda) \, d\lambda, \quad (4)$$

where $\mu$ is the Lambertian radiation order, $A_{\text{rec}}$ is the receiver aperture area, $\theta$ and $\psi$ denote the angles of transmission and reception, and $D$ is the distance between the transmitter and receiver. For the receiver, $G_{\text{am}}$ represents the amplifier gain, and $\kappa(\lambda)$ denotes the photodetector responsivity over the wavelength range.

As a result, through the combined OWC link model $H_{\text{owc}}(t)$, even under identical transmission conditions, the signal emitted by the LED will exhibit deviations from the original transmitted signal $s_{\text{tx}}(t)$, primarily in amplitude scaling and phase shift. These deviations, which vary from one LED to another, are then reflected in the received signal $s_{\text{rx}}(t)$ and can be exploited for LED identification.

Specifically, the amplitude variations in both the time domain and frequency domain can be expressed as follows.

**1) Time Domain Amplitude Variation:** The amplitude of the received signal $s_{\text{rx}}(t)$ in the time domain can be written as:

$$|s_{\text{rx}}(t)| = |h_{\text{led}}(t)| \otimes |h_{\text{channel}}(t)| \otimes |h_{\text{rx}}(t)| \otimes |s_{\text{tx}}(t)|, \quad (5)$$

where the LED response $h_{\text{led}}(t)$ introduces variations in the amplitude due to device-specific characteristics. In contrast, $h_{\text{channel}}(t)$ and $h_{\text{rx}}(t)$ remain constant under fixed conditions.

**2) Frequency Domain Amplitude Variation:** The amplitude of the received signal $S_{\text{rx}}(j\omega)$ in the frequency domain can be expressed as:

$$|S_{\text{rx}}(j\omega)| = |H_{\text{led}}(j\omega)| \cdot |H_{\text{channel}}(j\omega)| \cdot |H_{\text{rx}}(j\omega)| \cdot |S_{\text{tx}}(j\omega)|. \quad (6)$$

where, $H_{\text{led}}(j\omega)$ represents the LED frequency response,

which induces unique amplitude variations due to manufacturing tolerances and intrinsic differences among LEDs [20].

From the above analysis, it can be observed that the quantities influencing the amplitude, both in the time domain and frequency domain, are primarily affected by the LED response. Under fixed scenarios where the channel response $h_{\text{channel}}(t)$ and the receiver response $h_{\text{rx}}(t)$ remain unchanged, the amplitude variations observed in the received signal $s_{\text{rx}}(t)$ or $S_{\text{rx}}(j\omega)$ are solely attributed to the differences in LED hardware characteristics.

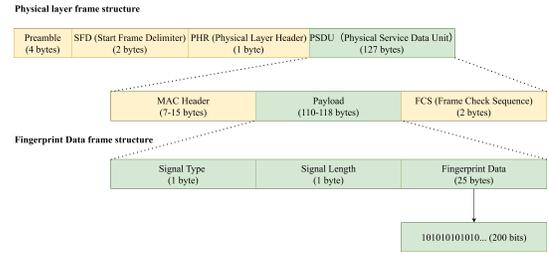

Fig. 4: Enhanced Physical Layer Frame and OFP Data Frame Structure.

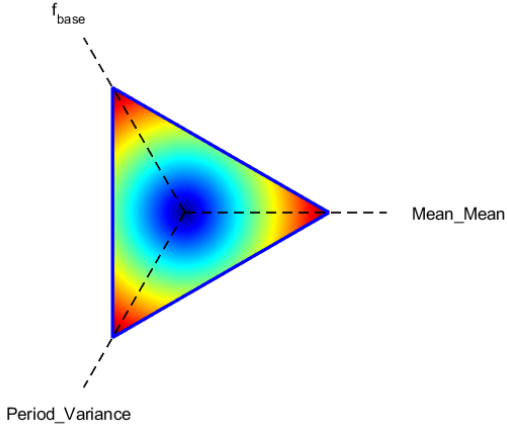

Fig. 3: Overall feature distribution of normalized OFP features.

Based on these findings, amplitude-based features, including the mean and variance of time-domain signals and the baseband amplitude in the frequency domain, are selected as OFP feature vectors. After normalization, these features are visualized to highlight device-specific distinctions. Fig. 3 illustrates the distinctiveness of these amplitude-based features. It demonstrates the capability of the proposed OFP method to differentiate LED devices reliably, forming the basis for robust device identification in LiFi networks.

### B. OFP date integrate in LiFi

To integrate the OFP mechanism into the LiFi network, we propose an enhanced data frame structure based on the IEEE 802.15.7 LiFi protocol. This protocol, designed for OWC, defines the physical and Media Access Control (MAC) layer frame structures for efficient data transmission and reception. The proposed OFP data frame incorporates fingerprint information into the Physical Service Data Unit (PSDU) payload section without introducing significant overhead.

As illustrated in Fig. 4, the data frame consists of two main components: the **Physical Layer Frame** and the embedded **Fingerprint Data Frame**.

*1) Physical Layer Frame::* The physical layer frame follows the IEEE 802.15.7 standard and consists of:

- **Preamble (4 bytes):** Synchronization between transmitter and receiver.
- **SFD (2 bytes):** Indicates the start of the frame.
- **PHR (1 byte):** Contains control information, such as frame length.
- **PSDU (127 bytes):** Carries the MAC header, payload, and FCS. A portion of the payload is allocated for fingerprint data.

*2) Fingerprint Data Frame::* Embedded within the PSDU payload, the fingerprint data frame includes:

- **Signal Type (1 byte):** Specifies the modulation scheme, e.g., On-Off Keying (OOK).
- **Signal Length (1 byte):** Indicates the size of the fingerprint data.
- **Fingerprint Data (25 bytes):** Contains the 200-bit binary sequence for extracting the OFP fingerprint.

This design ensures compatibility with IEEE 802.15.7 while enabling real-time identification of transmitting devices. Moreover, the enhanced data frame structure provides scalability for practical deployments, ensuring efficient transmission without introducing significant overhead.

### C. OFP Extraction Mechanism

The OFP extraction mechanism identifies LED-specific features embedded in transmitted signals, leveraging hardware-induced variations. The process consists of two main phases: **Transmission** and **Reception**.

*1) Transmission Process::* The OFP data frame is embedded within the PSDU payload and transmitted through the OWC channel. The process involves:

- Construction of the fingerprint data frame, which includes signal type, length, and fingerprint data.
- Modulation of the enhanced frame, e.g., using On-Off Keying (OOK).
- Signal transmission through the Tx chain (power amplifier, LED driver, and LED), where device-specific variations introduce unique distortions.

*2) Reception Process::* At the receiver, the transmitted signal undergoes preprocessing and feature extraction to recover fingerprint information:

- Frame synchronization and isolation of the fingerprint data.
- Extraction of key features (e.g., amplitude variations, rise/fall times, and frequency characteristics) reflecting LED-specific impairments.
- Formation of feature vectors for device identification.

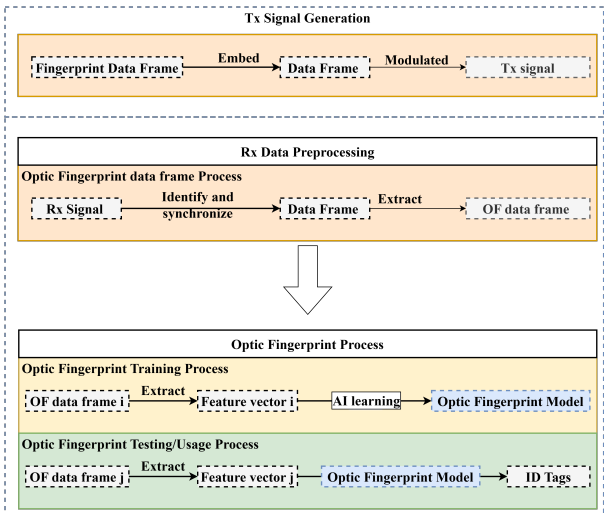

Fig. 5: OFP Extraction Mechanism: Transmission and Reception Processes.

*3) Training and Testing/Usage Phases::* The OFP extraction mechanism operates in two phases:

1) **Training Phase:** Feature vectors extracted from multiple OFP data frames are used to train a machine learning model (e.g., K-Nearest Neighbors(KNN), Support Vector Machine(SVM)).
2) **Testing/Usage Phase:** New feature vectors are matched against the trained model to authenticate devices and generate unique identifiers.

The proposed extraction mechanism, combined with the enhanced data frame structure, ensures efficient integration of OFP identification into LiFi networks. Leveraging hardware-induced variations enables real-time device identification while maintaining compatibility with existing protocols.

## III. Experiment and Results

To evaluate the performance of the proposed OFP mechanism, this study combines experimental data collected under controlled conditions with simulations that introduce channel noise and interference. Specifically, the experimental dataset is obtained at a short distance (3 cm) to eliminate channel effects, allowing the analysis to focus on the intrinsic variations caused by LED hardware. LiFi-specific channel characteristics, such as noise and interference, are subsequently modeled and introduced through simulations. This combined approach enables a comprehensive evaluation of the OFP extraction mechanism under ideal and practical conditions.

### A. Experiment Setup and Parameters

*1) Experiment Setup:* The experimental testbed for collecting OFP data is shown in Fig. 6. A 200 kHz square wave (10 V peak-to-peak) is generated by a waveform generator (Keysight 33500B) and transmitted via an OWC link. The optical signals are captured using a photodetector (New Focus Model 1601), powered by a power supply (Keysight E36234A), and recorded with an oscilloscope (Keysight MSOX6004A). The OFP database contains 39 LED samples (Cree XTE, Cree XPE2, LUMILEDS REBEL-LXMA, LG3535). Each sample is tested at 30 cm, chosen to minimize channel variations and isolate LED-specific impairments. While suitable for initial validation, future studies should assess OFP at longer ranges (e.g., 2–3 meters) for practical LiFi use.

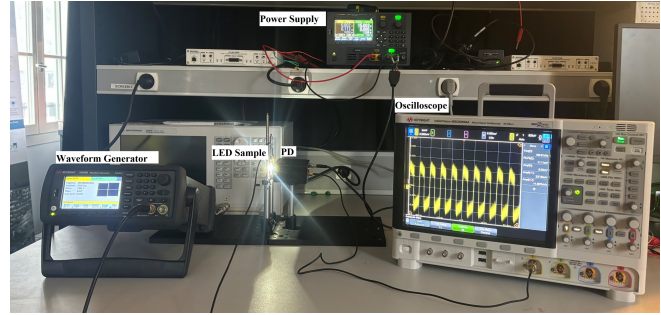

Fig. 6: Experimental Testbed for OF Data Collection.

*2) Feature Extraction:* The collected signal data is preprocessed to remove noise and align frames. Three amplitude-based features are extracted as follows:

- **Mean Amplitude:** Captures the average signal strength in the time domain.
- **Variance:** Quantifies the signal's dispersion in the time domain.
- **Baseband Amplitude:** Represents the signal's fundamental strength in the frequency domain.

These features are normalized to ensure consistency and comparability across different LED samples. The normalized feature vectors serve as the input for classification and analysis, enabling robust device identification.

*3) Simulation Parameters:* The robustness of the proposed OFP mechanism is evaluated using the following simulation parameters:

- **Channel Model:** Line-of-Sight (LOS)
- **Transmission Distance:** 5, 10, 15, 20, 25, and 30 cm
- **SNR Levels:** 10, 15, 20, 25, and 30 dB
- **Modulation:** On-Off Keying (OOK) at 200 kHz

### B. Results and Analysis

*1) OFP Extraction Results in Laboratory Environment:* The laboratory environment enabled the extraction of OFP features from 39 LEDs under controlled conditions. The features were normalized and visualized to evaluate their ability to differentiate devices. Fig. 7 illustrates the normalized feature distributions of all LED samples in a triangular coordinate system. Each triangle represents a unique LED, with the three vertices corresponding to the selected amplitude-based features: mean amplitude, variance, and baseband amplitude. These distributions confirm that the selected features effectively capture hardware-specific differences across LED devices, forming distinct clusters.

In addition to visual differentiation, the classification performance of the extracted features was evaluated using three classifiers: KNN, EnsembleKNN, and SVM_linear. Fig. 8 compares the test accuracy achieved using individual features (**f_base**, **Mean_Mean**, **Period_Variance**) and the combined feature set (**FixedFeatures**). The results show that the combined feature set consistently outperformed individual features across all classifiers, with SVM_linear achieving the highest accuracy at 91.08%.

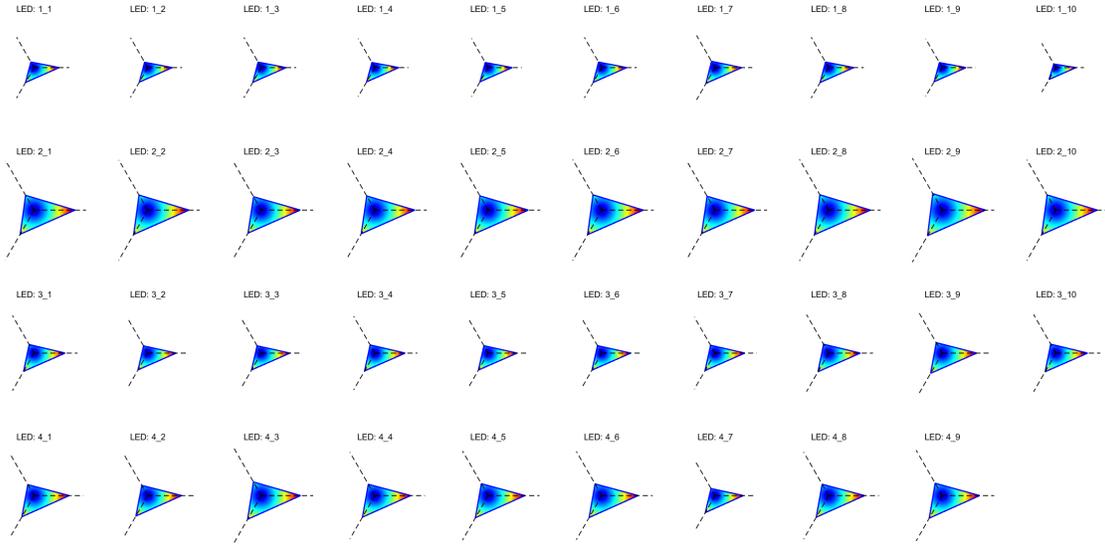

Fig. 7: Normalized feature distributions of 39 LEDs in a triangular coordinate system. Each triangle represents a unique LED, with the vertices corresponding to the selected amplitude-based features.

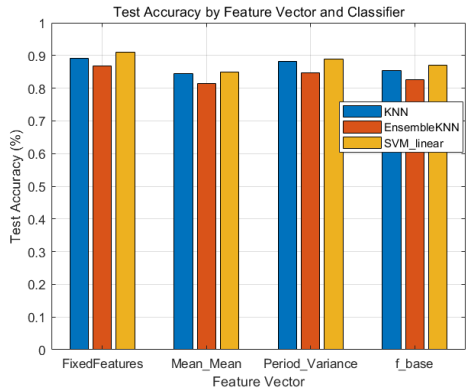

Fig. 8: Classification accuracy using different feature vectors (**f_base**, **Mean_Mean**, **Period_Variance**) and classifiers (**KNN**, **EnsembleKNN**, **SVM_linear**).

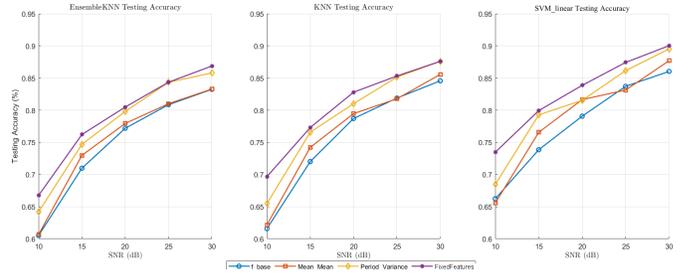

Fig. 9: Classification accuracy under noisy conditions for EnsembleKNN, KNN, and SVM_linear across different SNR levels.

*2) OFP extraction simulation result in different noise environment:* To assess the robustness of the OFP mechanism under noisy conditions, Gaussian noise was introduced to the dataset, simulating SNR levels ranging from 10 dB to 30 dB. Fig. 9 shows the classification accuracy for different classifiers under noisy conditions. The results indicate that the combined feature set maintains strong performance, achieving over 90.03% accuracy at 30 dB and above 70% at 10 dB. EnsembleKNN demonstrated the highest accuracy at lower SNR levels, while SVM_linear consistently outperformed other classifiers at higher SNRs. These findings confirm the robustness of the OFP mechanism in handling noise-induced signal distortions.

*3) OFP extraction simulation result in different distance environment:* The impact of transmission distance on the OFP mechanism was evaluated by simulating distances from 5 cm to 30 cm and SNR levels between 10 dB and 30 dB. Fig. 10 shows the classification accuracy across these conditions. Accuracy improves with increasing SNR, peaking at 90.36% at 30 dB. At 10 dB, accuracy remains above 70% across all distances. Under high SNR (30 dB), performance differences between distances are minimal, highlighting the robustness of the OFP mechanism to distance-related attenuation. However, at lower SNR levels, accuracy variability increases, reflecting a stronger dependency on SNR.

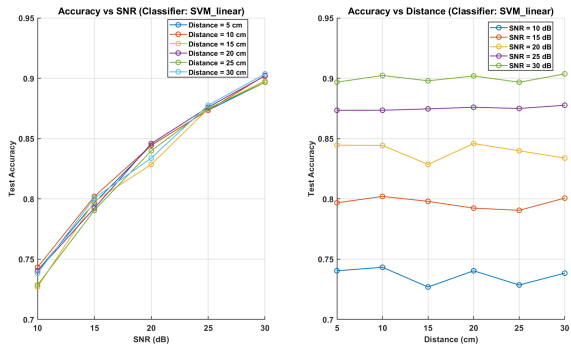

Fig. 10: Classification accuracy across SNR levels and transmission distances using the SVM_linear classifier.

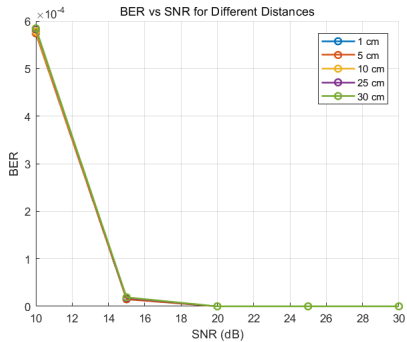

Fig. 11: BER performance under varying SNR levels and distances.

As shown in Fig. 11, the BER consistently meets the IEEE 802.15.7 standard [4], dropping below $10^{-3}$ at SNR levels of 10 dB or higher. These results confirm the robustness of the OFP mechanism against both channel noise and distance variations, validating its practicality for real-world applications.

## IV. Conclusion

This paper proposed a novel OFP framework to enhance the security of LiFi networks by leveraging LED hardware variations as unique identifiers. The framework extracts amplitude-based features—mean amplitude, baseband amplitude, and variance—and integrates them into an IEEE 802.15.7-compatible frame structure, enabling real-time device authentication. While experimental validation demonstrates promising results, future work should focus on evaluating environmental robustness, extending the communication range, and benchmarking OFP against other fingerprinting methods to ensure practical applicability in large-scale LiFi deployments.